\documentclass[a4paper,12pt]{article}
\usepackage{graphicx}
\usepackage{amsmath}
\usepackage{color}

\begin{document}

\begin{titlepage}
\vspace*{-3.5cm}
\vspace*{0.1cm} \rightline{KEK-CP-303,  NIKHEF 14-005}
\vspace*{0.1cm}\rightline{TTP14-004, SFB/CPP-14-06}
\vspace*{1.cm}
\begin{center}
{\Large{\bf Full $\mathcal{O}(\alpha)$ electroweak radiative corrections 
to $e^+e^- \rightarrow e^+e^-  \gamma$ at the ILC with GRACE-Loop}}\\

\vspace*{0.5cm}
  P.H. Khiem$^{A,B}$, Y. Kurihara$^{A}$, J. Fujimoto$^{A}$, T. Ishikawa$^{A}$,\\
 T. Kaneko$^{A}$, K. Kato$^C$, N. Nakazawa$^{C}$, Y. Shimizu$^A$, \\
 T. Ueda$^{D}$, J.A.M. Vermaseren$^E$, Y. Yasui$^{F}$\\
 
 \vspace*{1cm}

\textit{ $^{A)}$KEK, Oho 1-1, Tsukuba, Ibaraki 305-0801, Japan.\\ 
  $^{B)}$SOKENDAI University, Shonan Village, Hayama, Kanagawa 240-0193 Japan. \\ 
  $^{C)}$Kogakuin University, Shinjuku, Tokyo 163-8677, Japan.\\ 
  $^{D)}$Karlsruhe Institute of Technology (KIT), D-76128 Karlsruhe, Germany. \\
  $^{E)}$Nikhef, Science Park 105, 1098 XG Amsterdam, The Netherlands.\\ 
  $^{F)}$Tokyo Management College, Ichikawa, Chiba 272-0001, Japan.}
\end{center}
 \abstract{By using the GRACE-Loop system, we calculate the full 
$\mathcal{O}(\alpha)$ electroweak radiative corrections to the process 
$e^+e^- \rightarrow e^+e^-  \gamma$, which is important for future 
investigations at the International Linear Collider (ILC).  With the 
GRACE-Loop system, the calculations are checked numerically by three 
consistency tests: ultraviolet finiteness, infrared finiteness, and 
gauge-parameter independence. The results show good numerical stability 
when quadruple precision is used. In the phenomenological results, we find 
that the electroweak corrections to the total cross section range from 
$\sim -4\%$ to $\sim -21\%$ when $\sqrt{s}$ varies from $250$ GeV to $1$ 
TeV. The corrections also significantly affect the differential cross 
sections, which are a function of the invariant masses and angles and the 
final-particle energies. Such corrections will play an important role for 
the high-precision program at the ILC.}
\end{titlepage}

\section{Introduction}

The main goals of the International Linear Collider (ILC) are not only to 
precisely measure the properties of the Higgs particle, the top quark, and 
vector boson interactions but also to search for physics beyond the 
Standard Model. The high-precision measurements are expected to have a 
typical statistical error of less than $0.1\%$. This requires a very 
precise determination of the luminosity.

At the ILC, the integrated luminosity is 
measured~\cite{Bozovic-Jelisavcic:2013aca} by counting Bhabha events and 
comparing the result with the corresponding theoretical cross section:
\begin{eqnarray}
\int dt \; \mathcal{L} = \dfrac{N_{\mathrm{events}}-N_{\mathrm{bgk}}}
{\epsilon \;\cdot\sigma_{\mathrm{theory}}}.
\end{eqnarray}
In this formula $N_{\mathrm{events}} (N_{\mathrm{bgk}})$ is the number of 
the observed Bhabha events (the estimated background events). 
$\sigma_{\mathrm{theory}}$ is the Bhabha scattering cross section, which is 
calculated by using the perturbation theory. $\epsilon$ is the total 
selection efficiency for the events and $\int dt \; \mathcal{L}$ is the 
integrated luminosity.

A precise calculation of Bhabha scattering is important for a 
high-luminosity measurement, because the determination of all other cross 
sections depend on it. Thus, the one-loop electroweak corrections to Bhabha 
scattering are of considerable interest to many researchers. The full 
one-loop electroweak corrections to the $e^+e^- \rightarrow e^+e^-$ 
reaction were calculated many years ago in
Refs~\cite{tobimatsu1, tobimatsu2, Bohm:1984yt,Bohm:1986fg} and confirmed 
independently in Refs~\cite{Berends:1987jm, Fleischer:2006ht}. 
The corrections contribute significantly to the total cross section; 
about $\mathcal{O}(10\%)$ at high energy.

It is clear that the high-precision program at the ILC must consider the 
two-loop electroweak corrections to Bhabha scattering; many researchers 
have worked at these calculations for many years. However, the calculations 
were mostly performed at the level of two-loop QED corrections. To date, 
full two-loop electroweak corrections are not available. We refer here to 
several typical papers for two-loop QED calculations. A two-loop photonic 
correction to this process was calculated in Refs~\cite{Penin:2005kf,
Penin:2005eh}. In addition, two-loop QED corrections that maintain the 
electron mass in the squared amplitude are presented in 
Ref~\cite{Bonciani:2004gi}. In a later publication, the same group included 
the soft-photon-emission contribution to the differential cross section, as 
presented in Ref~\cite{Bonciani:2004qt}. We also like to mention the 
calculation of two-loop QED corrections to the Bhabha process which 
involves vacuum polarization by heavy fermions of arbitrary mass in Refs 
\cite{Bonciani:2007eh, Bonciani:2008ep}, two-loop QED corrections related to 
virtual hadronic and leptonic contributions to Bhabha scattering
also performed in Refs~\cite{Actis:2007fs,Actis:2008br}. Moreover, 
an approximation of the two-loop electroweak corrections to Bhabha scattering 
was computed in Ref~\cite{Penin:2011aa}. In this calculation, the authors 
considered the dominant logarithmically enhanced two-loop electroweak corrections
to the differential cross section in the high-energy limit and at large scattering 
angles.

The perspectives of the present calculation are as follow: To correct the 
number of Bhabha events, a precise evaluation of its background is 
required. Experiments may misidentify $e^+e^-\gamma$ as $e^+e^-$ events 
because (i) the photon is a hard bremsstrahlung photon that can escape the 
detector, (ii) the photon is a soft bremsstrahlung photon that has a small 
opening angle with respect to the final electron (or positron), or (iii) 
the photon is emitted in parallel to the beam axis. With these 
misidentifications, the process $e^+e^- \rightarrow e^+e^-\gamma$ is one 
channel that contributes significantly to the background of Bhabha events. 
Hence the precise calculation of the process is of great importance. 
Furthermore, in the framework of calculating the full two-loop corrections 
to Bhabha scattering, one-loop electroweak corrections to $e^+e^- 
\rightarrow e^+e^-\gamma$ with a soft bremsstrahlung photon are necessary; 
they should cancel against the infrared divergences which appear at the 
level of two-loop corrections to Bhabha scattering. Last but by no means 
least, the process will be a good candidate for luminosity measurements at 
the ILC, provided these theoretical calculations are well under control.

We refer to a few additional papers that should be mentioned. The 
lowest-order calculation of the soft-bremsstrahlung process is reported in 
Ref. \cite{Tobimatsu:2001kb}. Moreover, the one-loop QED corrections to the 
hard-bremsstrahlung process $e^+e^- \rightarrow e^+e^-  \gamma$ is 
available in Ref~\cite{Actis:2009uq}. An analytical calculation of one-loop 
QED corrections to the process $e^+e^- \rightarrow e^+e^-  \gamma$ is also 
calculated in Ref~\cite{Igarashi}.

To achieve our eventual target, the calculation of two-loop corrections to 
Bhabha scattering, several steps are involved, the first of which is to 
consider the process as a candidate for luminosity measurements at the ILC, 
because it provides a useful framework for our final objective. This is 
what we present in this paper. In particular, we focus on studying the 
impact of electroweak corrections to the total cross section and to the 
relevant distributions such as the cross sections that are functions of the 
invariant masses, energies, and angles of the final state particles. We will 
incorporate the soft photon bremsstrahlung and subsequently the two-loop 
corrections to Bhabha scattering in future publications.

The layout of the paper is as follows: In Section 2, we present a short 
introduction to the GRACE-Loop system and the numerical tests of the 
calculation. In Section 3, we present the phenomenological results of the 
calculation. Conclusions and plans for future work are presented in Section 
4.


\section{GRACE Loop and the $e^+e^- \rightarrow e^+e^-  \gamma$ process }

\subsection{GRACE Loop}
GRACE Loop is a generic program that automates the calculation of 
high-energy physics processes at the one-loop level. The program is 
described in detail in Ref~\cite{Belanger:2003sd}, where a variety of 
electroweak processes with two particles in the final state are presented 
and compared with other papers. The GRACE-Loop system was also used to 
calculate processes with three particles in the final state, such as 
$e^+e^- \rightarrow ZHH$~\cite{Belanger:2003ya}, $e^+e^- \rightarrow t 
\bar{t} H$~\cite{Belanger:2003nm}, and $e^+e^- \rightarrow \nu \bar{\nu} 
H$~\cite{Belanger:2002me}. These calculations were performed independently 
by several groups; for example, the processes $e^+e^- \rightarrow 
ZHH$~\cite{Zhang:2003jy}, $e^+e^- \rightarrow t \bar{t} 
H$~\cite{You:2003zq, Denner:2003ri, Denner:2003zp}, and $e^+e^- \rightarrow 
\nu \bar{\nu} H$~\cite{Denner:2003yg, Denner:2003iy}. In addition, the 
$e^+e^- \rightarrow \nu_{\mu} \bar{\nu}_{\mu} HH$~\cite{Kato:2005iw} 
reaction was calculated by using the GRACE-Loop system.

In the GRACE-Loop system, the renormalization is performed with the 
on-shell renormalization condition of the Kyoto scheme, as described in 
Ref~\cite{kyotorc}. Ultraviolet (UV) divergences are regulated by 
dimensional regularization, and infrared (IR) divergences are 
regularized by giving the photon an infinitesimal mass $\lambda$. In the 
current version there are no soft external gluons.

The GRACE-Loop system uses the symbolic-manipulation package FORM 
~\cite{form, form4.0} to handle all Dirac and tensor algebra in $n$ 
dimensions. It symbolically reduces all tensor one-loop integrals to scalar 
integrals. Eventually, the amplitude of the given processes will be written 
in terms of FORTRAN subroutines on a diagram-by-diagram basis.

Ref~\cite{Belanger:2003sd} describes the method used by the GRACE-Loop 
system to reduce tensor one-loop five- and six-point functions to one-loop 
four-point functions. The tensor one-, two-, three-, and four-point 
functions are then reduced to scalar one-loop integrals that are 
numerically evaluated by one of the FF~\cite{ff} or 
LoopTools~\cite{looptools} packages. 

The GRACE-Loop program uses so-called nonlinear gauge fixing terms 
\cite{nlg-generalised} in the Lagrangian, which are defined as
\begin{eqnarray}
{{\cal L}}_{GF}&=&-\dfrac{1}{\xi_W}
|(\partial_\mu\;-\;i e \tilde{\alpha} A_\mu\;-\;ig c_W
\tilde{\beta} Z_\mu) W^{\mu +} + \xi_W \dfrac{g}{2}(v
+\tilde{\delta} H +i \tilde{\kappa} \chi_3)\chi^{+}|^{2} \nonumber \\
& &\;-\dfrac{1}{2 \xi_Z} (\partial\cdot Z + \xi_Z \dfrac{g}{ 2 c_W}
(v+\tilde\varepsilon H) \chi_3)^2 \;-\dfrac{1}{2 \xi_A} (\partial \cdot A
)^2 \;.
\end{eqnarray}
We are working in the $R_{\xi}$-type gauges with the condition 
$\xi_{W}=\xi_{Z}=\xi_{A}=1$ (also called the 't Hooft-Feynman gauge) in 
which there is no longitudinal contribution to the gauge propagator. This 
choice not only has the advantage of making the expressions much simpler 
but also avoids unnecessarily large cancellations, high tensor ranks in 
one-loop integrals, and extra powers of momenta in the denominators, which 
cannot be handled by the FF and LoopTools packages. The implementation of 
nonlinear gauge-fixing terms provides a powerful tool to check the results 
in a consistent way. After all, the results must be independent of the 
nonlinear gauge parameters, as will be discussed in greater detail in 
subsection 2.2.

In its latest version, the GRACE Loop system can use the axial gauge in the 
projection operator for external photons. This resolves a problem 
with large numerical cancellations, which is very useful when calculating 
processes at small angle and energy cuts for the final-state particles. 
Moreover, it provides a useful tool to check the consistency of the 
results, which due to the Ward identities, are independent of the choice of 
the gauge. This method was applied to the process $e^+e^- \rightarrow t 
\bar{t} \gamma$ in Ref~\cite{Khiem:2012bp}, and we apply it here as well. 
For the integration steps, we use a parallel version of BASES~\cite{bases} 
with a message-passing interface~\cite{mpi} to reduce the calculation 
time.

\subsection{The $e^+e^- \rightarrow e^+e^- \gamma$ process } 

\begin{figure}[h]
 \begin{center}
 \includegraphics[width=6.3in, height=12cm]{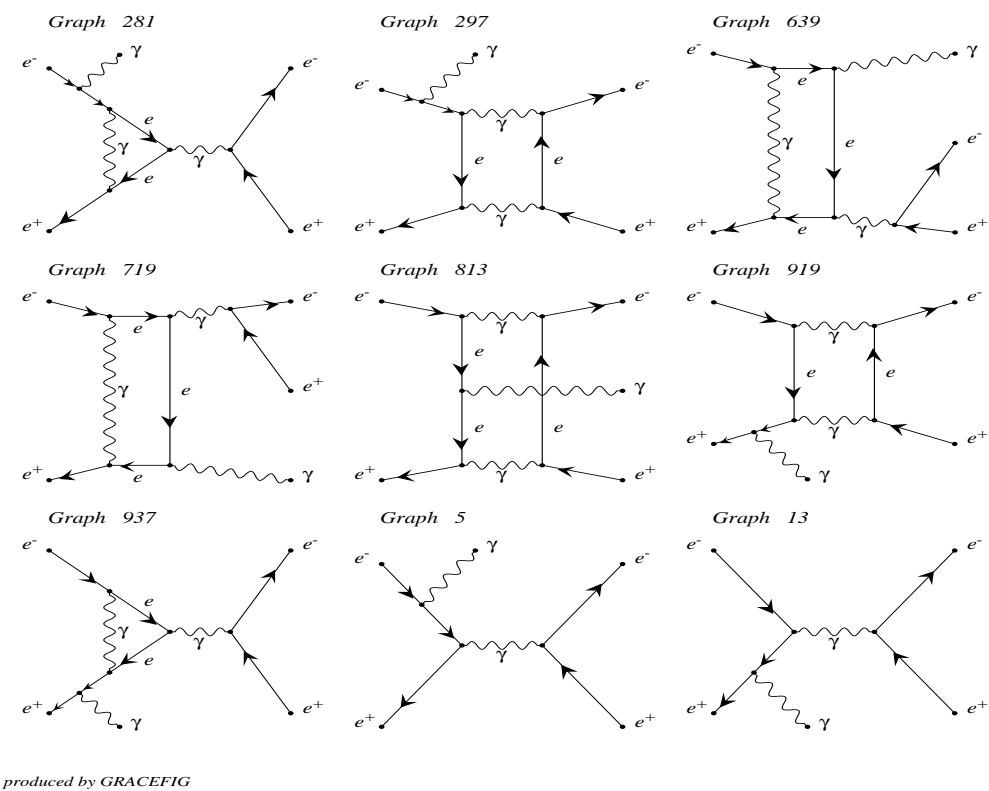}
 \end{center}
\caption{\label{feynmandiagrams} Typical Feynman diagrams for the reaction 
$e^+e^- \rightarrow e^+e^- \gamma$ as generated by the GRACE-Loop system.}
\end{figure}

The full set of Feynman diagrams with the nonlinear gauge fixing, as 
described in the previous section, consists of $32$ tree diagrams and  
$3456$ one-loop diagrams. This includes the counterterm diagrams. In 
Fig.~\ref{feynmandiagrams}, we show some selected diagrams.
For this calculation, we apply an axial gauge for the external photon by 
using the polarization sum of the photons as follows:
\begin{eqnarray}
\label{polarization}
\mathcal{P}(\lambda)= \sum\limits_{\lambda=0}^{3} 
\epsilon_{\lambda}^{\mu}(p)\epsilon_{\lambda}^{\mu}(p) \rightarrow -g^{\mu\nu}
 + \dfrac{n^{\mu}p^{\nu} +n^{\nu} p^{\mu} }{n\cdot p} - n^2 \dfrac{p^{\mu}p^{\nu}}{(n\cdot p)^2},
\end{eqnarray}
where $p^{\mu}$ and $\epsilon_{\lambda}^{\mu}$ correspond to the 
$4$-momentum and the polarization vector of the external photon 
respectively. The axial vector $n$ takes the form 
\begin{eqnarray}
\label{axial}
 n=(p^0, -\vec{p}).
\end{eqnarray}
With this choice, the third term in Eq.~(\ref{polarization}) vanishes, 
which means that we are working in the light-cone gauge for the photon. The 
advantage of using the axial gauge for the external photon is that the 
worst numerical cancellations between the diagrams are avoided.

Before running the Monte Carlo integration for the process, the 
calculation is checked numerically by three consistency tests. These are UV 
and IR finiteness and gauge-parameter independence. The general idea of 
these tests is now described.

The full $\mathcal{O}(\alpha)$ electroweak cross section considers the tree 
graphs and the full one-loop virtual corrections as well as the soft and 
hard bremsstrahlung contributions. In general, the total cross section in 
full one-loop electroweak radiative corrections is given by
\begin{eqnarray}
\label{tot}
 \sigma^{e^-e^+ \gamma_ H}_{\mathcal{O}(\alpha)} &=& \int d\sigma^{e^-e^+ \gamma_H} _{\textbf{\small {T}}}
 +\int d\sigma^{e^-e^+ \gamma_H}_{\textbf{\small V}}(C_{UV},
 \{\tilde{\alpha}, \tilde{\beta}, \tilde{\delta},\tilde{\epsilon},  \tilde{\kappa} \}, \lambda)
 \nonumber \\ &&
 + \int d\sigma^{e^-e^+ \gamma_H}_{\textbf{\small T}} \delta_{\textbf{\small soft}}(\lambda \leq E_{\gamma_S} <k_c)
 + \int d\sigma^{e^-e^+ \gamma_H \gamma_S}_{\textbf{\small H}}(E_{\gamma_S} \geq k_c).
  \end{eqnarray}
In this formula, $\sigma^{e^-e^+ \gamma_H}_{\textbf{\small T}} $ is the 
tree-level cross section, $\sigma^{e^-e^+ \gamma_H}_{\textbf{\small V}}$ is 
the cross section due to the interference between the one-loop (including 
counterterms) and the tree diagrams. The contribution must be independent 
of the UV-cutoff parameter ($C_{UV}$) and the nonlinear gauge parameters 
($\tilde{\alpha}, \tilde{\beta}, \tilde{\delta}, \tilde{\epsilon},  
\tilde{\kappa}$). Because of the way we regularize the IR divergences, 
$\sigma^{e^-e^+ \gamma_H}_{\textbf{\small V}}$ depends on the photon mass 
$\lambda$. This $\lambda$ dependence must cancel against the soft-photon 
contribution, which is the third term in Eq. (\ref{tot}). The soft-photon 
contribution can be factorised into a soft factor, which is calculated 
explicitly in Ref~\cite{supplement100}, and the cross section from the tree 
diagrams.

In Tables \ref{cuv}, \ref{gauge}, and \ref{lambda} in the appendix, we 
present the numerical results for the checks of UV finiteness, gauge 
invariance, and the IR finiteness for one random point in phase space, 
calculated with quadruple precision. The results are stable over a range of 
20 digits. The different precisions are due to the ways in which these 
parameters occur in the formulas: $C_{UV}$ occurs only linearly as an extra 
term, and the nonlinear gauge parameters occur as products in terms that 
are by themselves typically much larger than the remaining terms. The IR 
regulator $\lambda$ contributes mainly because of its appearance in the 
denominators and hence occurs inside logarithms. Consequently, the $C_{UV}$ 
checks show an agreement in more digits than the other checks.

Finally, we consider the contribution of the hard photon bremsstrahlung, 
$\sigma^{e^-e^+ \gamma_H \gamma_S}_{\textbf{\small H}}(k_c)$. This part is 
the process $e^+e^- \rightarrow e^- e^+ \gamma_H \gamma_S$ with an added 
hard bremsstrahlung photon. The process is generated by the tree-level 
version of the GRACE system~\cite{grace} with the phase space integration 
performed by BASES. By adding this contribution to the total cross section, 
the final results must be independent of the soft-photon cutoff energy 
$k_c$. Table \ref{kc} in the appendix shows the numerical result of the 
check of $k_c$ stability. By changing $k_c$ from $10^{-3}$ GeV to $0.1$ 
GeV, the results are consistent to an accuracy better than $0.04\%$ (this 
accuracy is better than that in each Monte Carlo integration). For the 
check of $k_c$ stability, note that we have two photons at the final state. 
One photon is the hard photon to which we apply an energy cut of 
$E_{\gamma_H}^{\mathrm{cut}} \geq 10$ GeV and an angle cut of 
$10^{\circ}\leq \theta_{\gamma_H}^{\mathrm{cut}} \leq 170^{\circ}$. The 
second photon is the soft photon whose energy is greater than $k_c$ and 
smaller than the energy of the first photon.

Having verified the stability of the results, we proceed to compute the 
physics of the process. Hereafter, we use $\lambda=10^{-21}$ GeV, 
$C_{UV}=0$, $k_c=10^{-3}$ GeV, and $\tilde{\alpha}= \tilde{\beta}= 
\tilde{\delta}= \tilde{\kappa}=\tilde\varepsilon=0$. To reduce the 
calculation time, we neglect the diagrams that contain the coupling of the 
Higgs boson to the electron and positron in the integration step because 
its contribution is less than the statistical error of the Monte Carlo 
integration.


\section{Results of the calculation}

We used the following input parameters for the calculation:
\begin{description}
\item[] The fine structure constant in the Thomson limit is 
$\alpha^{-1}=137.0359895$.
\item[] The mass of the Z boson is $M_Z=91.1876$ GeV and its decay width is 
$\Gamma_Z= 2.35$ GeV.
\item[] The mass of the Higgs boson is taken to be $M_H=126$ GeV.
\item[] In the on-shell renormalization scheme we like to take the mass of 
the $W$ boson as an input parameter. Because of the limited accuracy of the 
measured value, we take the value that is derived from the electroweak 
radiative corrections to the muon decay width ($\Delta 
r$)~\cite{Hioki:1995ex} with $G_{\mu}=1.16639 \times 10^{-5}$ GeV$^{-2}$. 
Therefore, $M_W$ is a function of $M_H$. This results in $M_W = 80.370$ GeV 
as explained in subsection \ref{EWcorrections}, corresponding to $\Delta 
r=2.49\%$.
\item[] For the lepton 
masses we take $m_e=0.51099891$ MeV, $m_{\mu}=105.658367$ MeV and 
$m_{\tau}=1776.82$ MeV.
\item[] For the quark masses, we take $m_u=63$ MeV, $m_d=63$ MeV, $m_c=1.5$ 
GeV, $m_s=94$ MeV, $m_t=173.5$ GeV, and $m_b=4.7$ GeV.
\end{description}

Because the process considered in this paper is a candidate for luminosity 
measurements, the full $\mathcal{O}(\alpha)$ electroweak corrections to 
$e^-e^+ \rightarrow e^-e^+\gamma$ are evaluated by applying cuts that are 
suitable for this purpose. For the final-state particles, we apply an 
energy cut $E^{\mathrm{cut}} \geq 10$ GeV and an angle cut $10^{\circ}\leq 
\theta^{\mathrm{cut}} \leq 170^{\circ}$ with respect to the beam axis. 
Moreover, to isolate the photon from the electron (or positron), we apply 
an opening angle cut between the photon and the $e^-$($e^+$) of 
$10^{\circ}$. Finally, to distinguish $e^- e^+ \gamma$ events from $\gamma 
\gamma$ events, we apply an angle cut of $10^{\circ}$ between the electron 
and the positron in the final state. 

The results for this case are presented in the following 
subsection. The two-loop corrections to the Bhabha-scattering calculation 
will be part of a future project.

\subsection{Total cross section and electroweak corrections}
\label{EWcorrections}

The total cross section is calculated by using Eq.~(\ref{tot}). The relative 
correction is then defined in the $\alpha$ scheme as
\begin{eqnarray}
 \delta_{EW} & = & K_{EW} - 1 \\
             & = & \dfrac{\sigma_{\mathcal{O}(\alpha)}}{\sigma_{\bf{tree}} }-1,
\end{eqnarray}
where the term $K_{EW}$ is the ratio of the full cross section up to one-loop 
radiative corrections to the cross section from tree-level contributions.

In the GRACE-Loop system, the QED corrections can be calculated separately 
by selecting individual QED diagrams and their counterterms. As expressed in 
the following equation, the total QED cross section is then normalized to 
the cross section of the full tree diagrams to extract the QED corrections:
\begin{eqnarray}
 \delta_{QED} & = &
    \dfrac{\sigma_{\bf{V+S+H}}^{\mathrm{\small{QED}}}}{\sigma_{\bf{tree}} }.
\end{eqnarray}
The next equation gives the genuine weak correction in the $\alpha$ scheme: 
\begin{eqnarray}
 \delta_{W} & = & \delta_{EW}-\delta_{QED}. 
\end{eqnarray}
Having subtracted the genuine weak corrections in the $\alpha$ scheme, one 
can express the correction in the $G_{\mu}$ scheme. This approach is also 
called the improved Born approximation, where the fine structure constant 
runs from the Thomson-limit condition to the $M_Z^2$ scale. Some of the 
high-order corrections are related to two-point functions, which are 
connected to light fermions and absorbed into the tree-level calculation. 
To obtain the corrections in this scheme, we subtract the universal weak 
correction obtained from $\Delta r$ as follows:~\footnote{The order of 
$\alpha$, which comes from the coupling of real photons to fermions, must 
be calculated under the conditions of the Thomson limit. The order 
$\alpha^2$ runs from the Thomson limit to the $M_Z^2$ scale. Overall, these 
considerations lead to the factor $2$ in Eq. (\ref{gmu}).}
\begin{eqnarray}
\label{gmu}
\delta^{G_{\mu}}_{W}= \delta_{W} - 2 \Delta r,
\end{eqnarray}
with $\Delta r=2.49\%$ for $M_H=126$ GeV. 

Table~\ref{cross-ew} shows the total cross section and the electroweak 
corrections as a function of $\sqrt{s}$. The center-of-mass energy ranges 
from 250 GeV (which is near the threshold of $M_H+M_Z$) to $1$ TeV.

We find that the electroweak (QED) corrections in the $\alpha$ scheme vary 
from $\sim-4\%$ ($\sim -5\%$) to $\sim-21\%$  ($\sim-17\%$) as $\sqrt{s}$ 
varies from $250$ GeV to $1$ TeV. The results given in Table \ref{cross-ew} 
show clearly that the QED corrections make the dominant contribution 
compared with the weak corrections. The weak corrections in the $G_{\mu}$ 
scheme vary from $\sim -4\%$ to $\sim -9\%$ as $\sqrt{s}$ varies from $250$ 
GeV to $1$ TeV. The weak corrections in the high-energy region are 
attributed to the enhancement contribution of the single Sudakov logarithm. 
Its contribution can be estimated as follows:
 \begin{eqnarray}
  \delta^{G_{\mu}}_{W} \sim -\frac{\alpha(M_Z^2)}{\pi \; \mathrm{sin}^2\theta_{\mathrm{W}}}
  \log(\frac{s}{M_Z^2})\sim \mathcal{O}(-10\%) \quad \mathrm{at}\quad \sqrt{s}=1 \; \mathrm{TeV}.
 \end{eqnarray}
It is clear that the corrections make a sizable contribution to the total 
cross section and cannot be ignored for the high-precision program at the 
ILC.
\begin{table}[htbp]
\centering
\begin{tabular}{ccccccccc } \hline 
$\sqrt{s}$ [GeV]      &    $\sigma_{\bf{T}}$ [pb]  
&    $\sigma_{\mathcal{O}(\alpha)}^{\tiny \mathrm{QED}}$  [pb]
&    $\sigma_{\mathcal{O}(\alpha)}$  [pb]
& $\delta_{\mathrm{QED}} [\%]$   
& $\delta_{\mathrm{EW}} [\%]$  
& $\delta_{\mathrm{W}} [\%]$
& $\delta_{\mathrm{W}}^{G_{\mu}} [\%]$ \\ \hline\hline
250  &	9.746   &9.269    &    9.317  &  -4.89  & -4.40    &0.49   &   -4.49 \\ \hline
350  & 	5.684   &5.244    &    5.254  &  -7.74  & -7.57    &0.17   &   -4.81\\ \hline
500  &	3.175   &2.839    &    2.811  & -10.58 & -11.47    &-0.89  &   -5.87\\ \hline
700  &	1.817   &1.564    &    1.534  & -13.92 & -15.58    &-1.66  &   -6.64\\ \hline
1000 &	1.001   &0.828    &    0.789  & -17.28 & -21.18    &-3.90  &   -8.88\\ \hline\hline
\end{tabular}
\caption{\label{cross-ew} The total cross section and the electroweak 
corrections as a function of the center-of-mass energy.}
\end{table}

\subsection{Relevant distributions}

We now generate the relevant distributions such as the cross sections, 
which are a function of the invariant masses, energies, and angles of the final 
state particles. In these distributions, the solid lines represent the 
tree-level calculation, and the points with error bars 
include the full radiative corrections. The left (right) figures show the 
given distributions at $\sqrt{s}=250$ GeV ($1$ TeV). The term $K_{EW}$ is 
also shown with these distributions to estimate the electroweak corrections 
to the differential cross sections.

\begin{figure}[htpb]
\begin{center}$
\begin{array}{cc}
\hspace*{-0.6cm}\includegraphics[width=8.54cm,height=8.65cm, angle=-90]{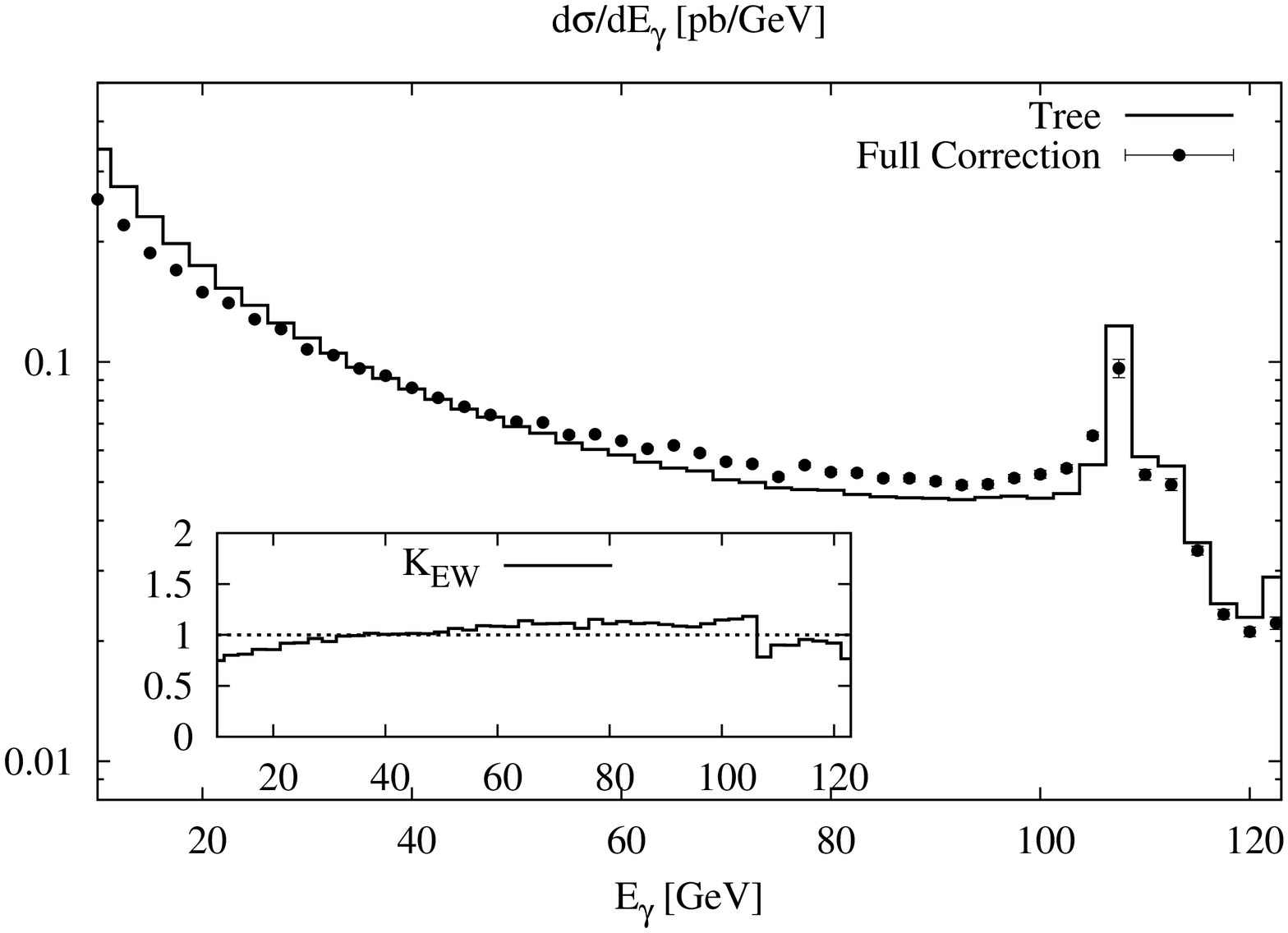} &
\hspace*{-1.0cm}\includegraphics[width=8.54cm,height=8.65cm, angle=-90]{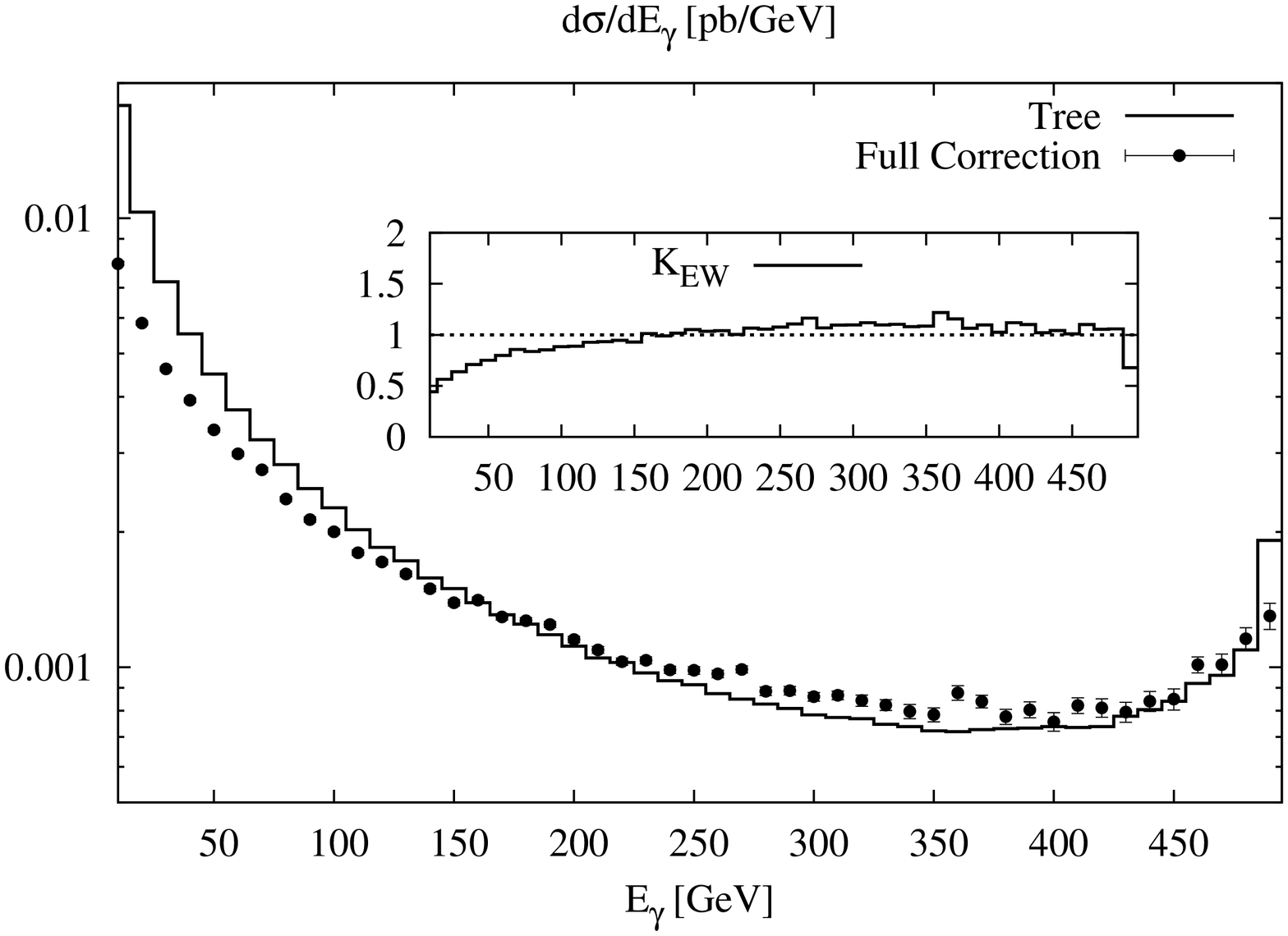}\\
\hspace{-0.27cm}\includegraphics[width=8.21cm,height=8.32cm, angle=-90]{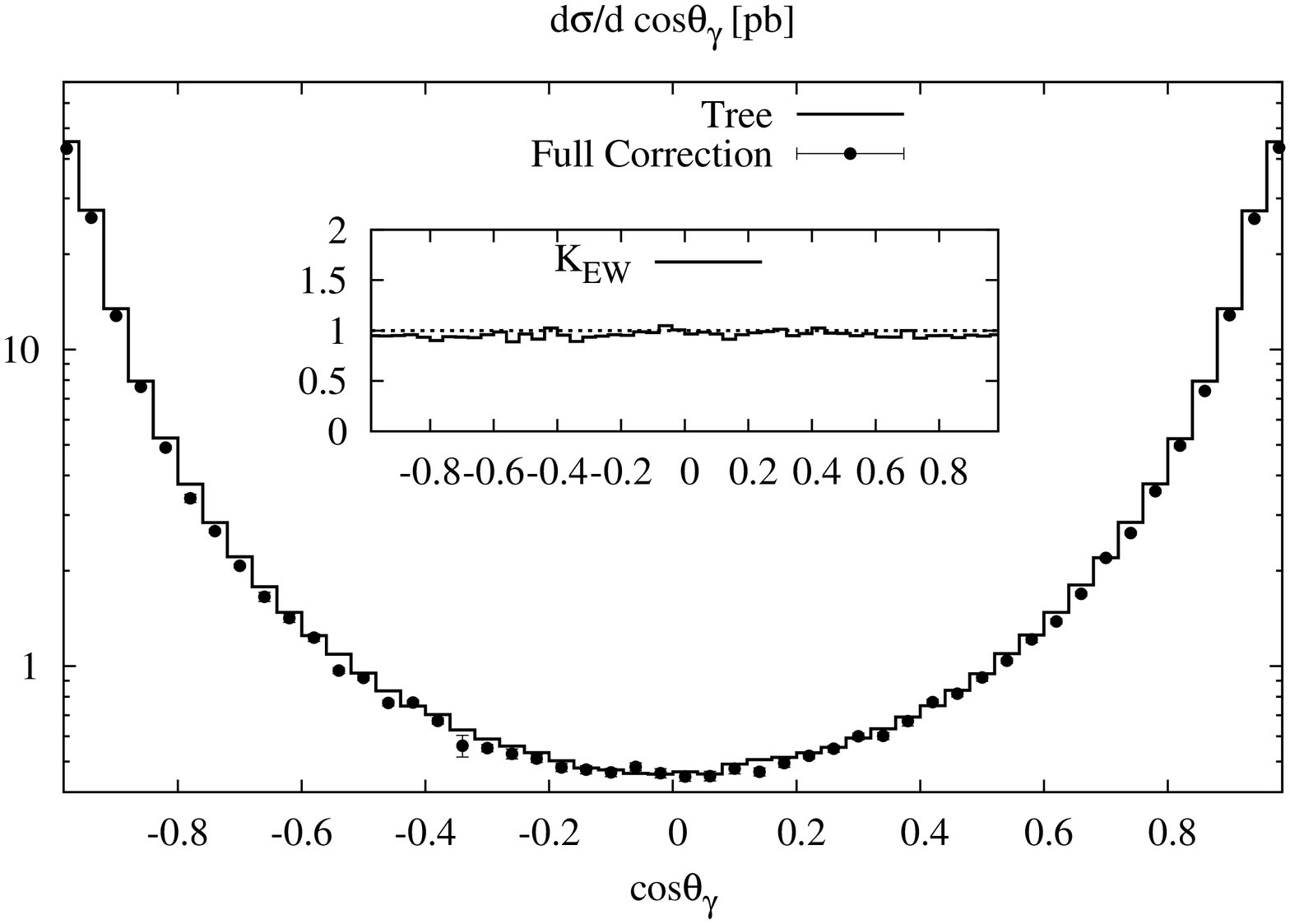} &
\hspace{-0.67cm}\includegraphics[width=8.21cm,height=8.32cm, angle=-90]{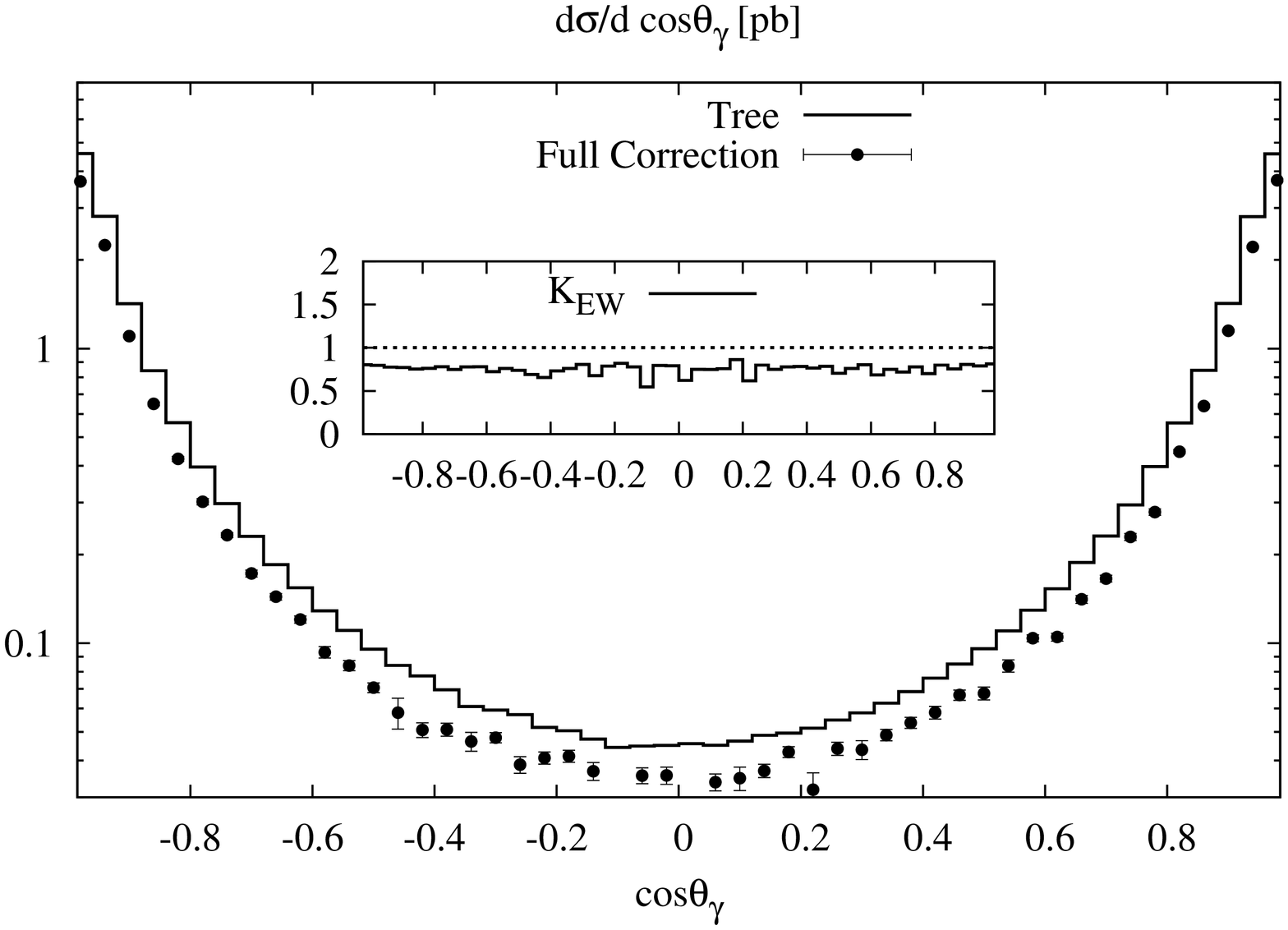}\\
\end{array}$
\end{center}
\caption{\label{photonenergy} Differential cross sections as a function 
of photon energy and cos$\theta_{\gamma}$ 
 at (left panel) $\sqrt{s}=250$ GeV and (right panel) $\sqrt{s}=1$ TeV.}
\end{figure}

Figure~\ref{photonenergy} presents the cross-section distributions as a 
function of the photon energy for $\sqrt{s}=250$ GeV and $\sqrt{s}=1$ TeV. 
Overall, the cross section decreases with increasing photon energy. At 
$\sqrt{s} = 250$ GeV, two peaks appear, one at 
$E_{\gamma}=\dfrac{s-M_Z^2}{2\sqrt{s}}$ and one at $\dfrac{\sqrt{s}}{2}$. 
The first peak corresponds to the photon energy recoiling against an 
on-shell $Z$ boson, and the right peak corresponds to the photon energy 
recoiling against a virtual photon that creates a small-mass 
electron-positron pair. Due to the high energy the peaks overlap within our 
resolution at $\sqrt{s}=1$ TeV. The distributions also clearly show that 
the radiative corrections make a sizeable impact and are important for the 
luminosity monitor at the ILC. The lower part of Fig.~\ref{photonenergy} 
shows the angular distributions of the photon at $\sqrt{s}=250$ GeV and 
$\sqrt{s}=1$ TeV. The cross section is symmetric with respect to 
cos$\theta_{\gamma}$. The radiative corrections make a more significant 
contribution at $\sqrt{s}=1$ TeV compared with their contribution at $250$ 
GeV center-of-mass energy.

\begin{figure}[hbtp]
\begin{center}$
\begin{array}{cc}
\hspace*{-0.6cm}\includegraphics[width=8.54cm,height=8.65cm, angle=-90]{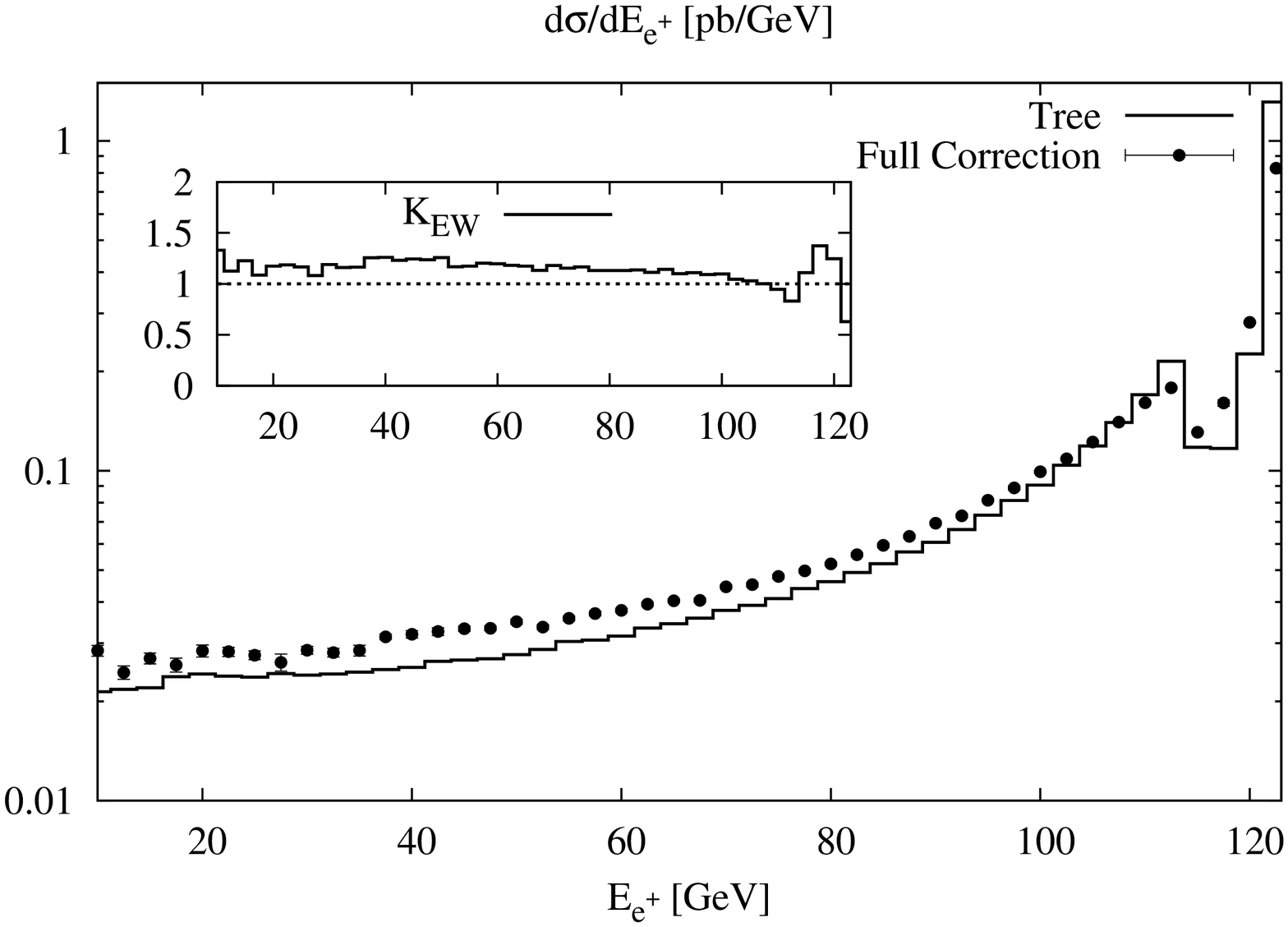} &
\hspace*{-1.0cm}\includegraphics[width=8.54cm,height=8.65cm, angle=-90]{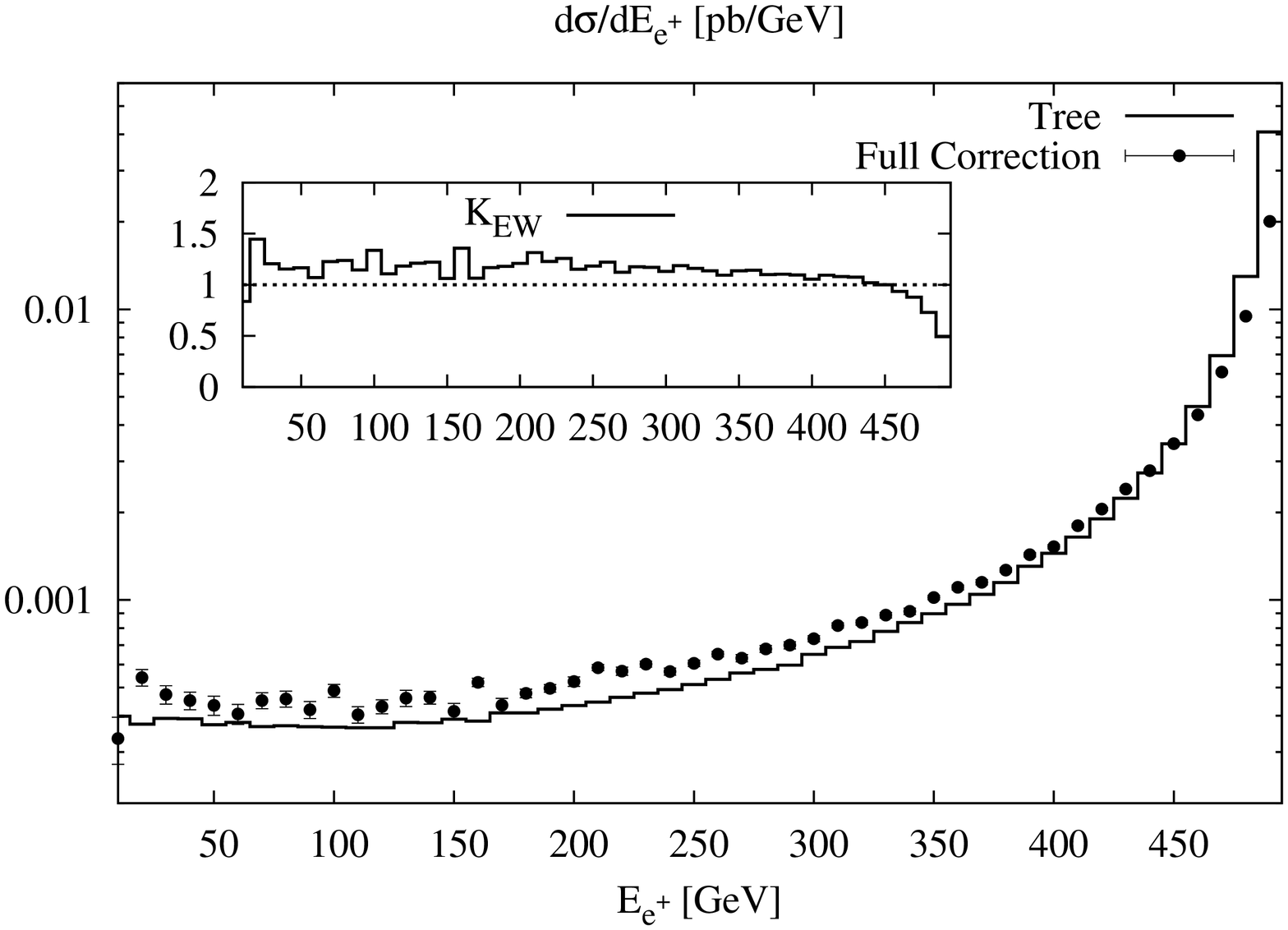}\\
\hspace{-0.27cm}\includegraphics[width=8.21cm,height=8.32cm, angle=-90]{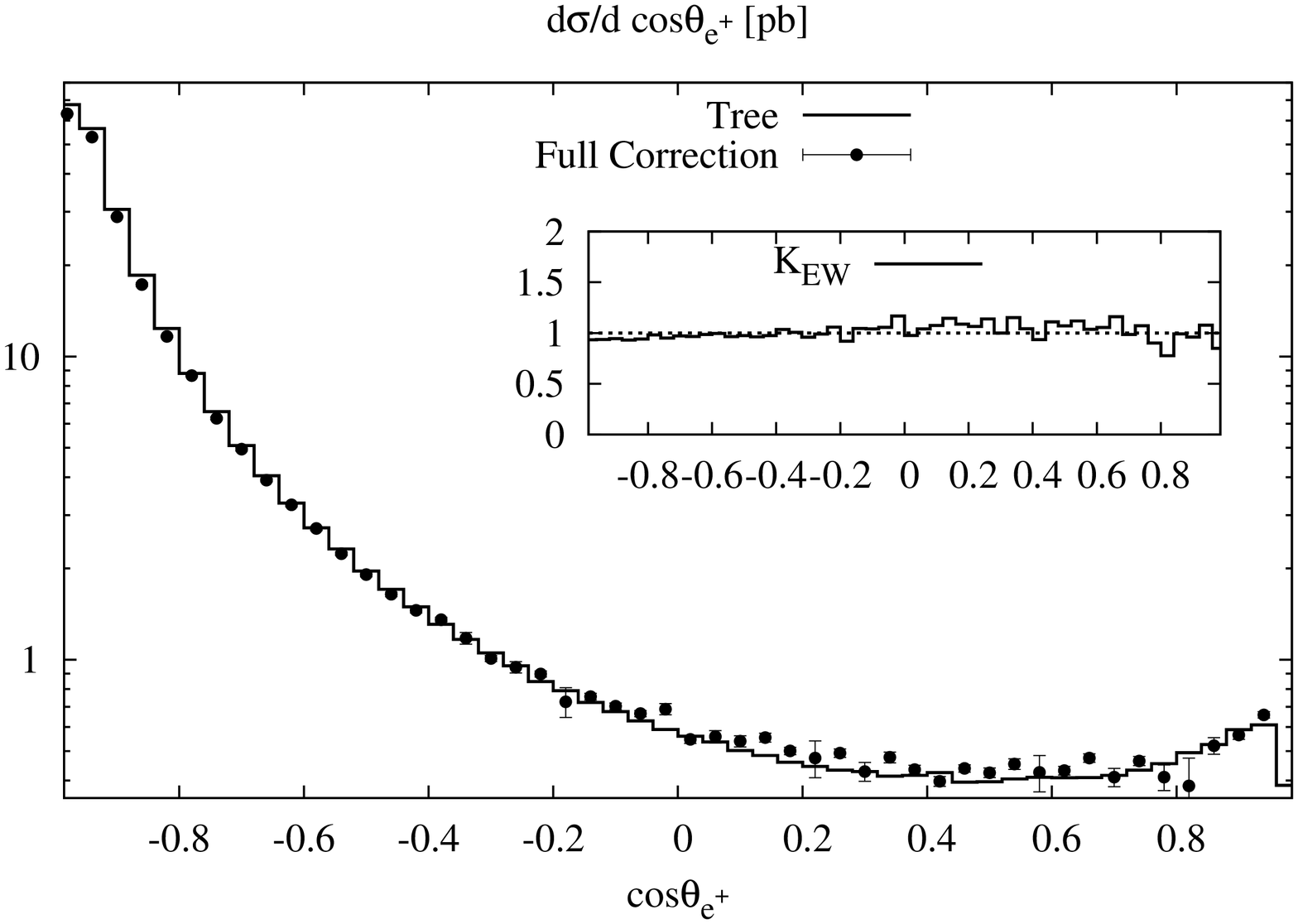} &
\hspace{-0.67cm}\includegraphics[width=8.21cm,height=8.32cm, angle=-90]{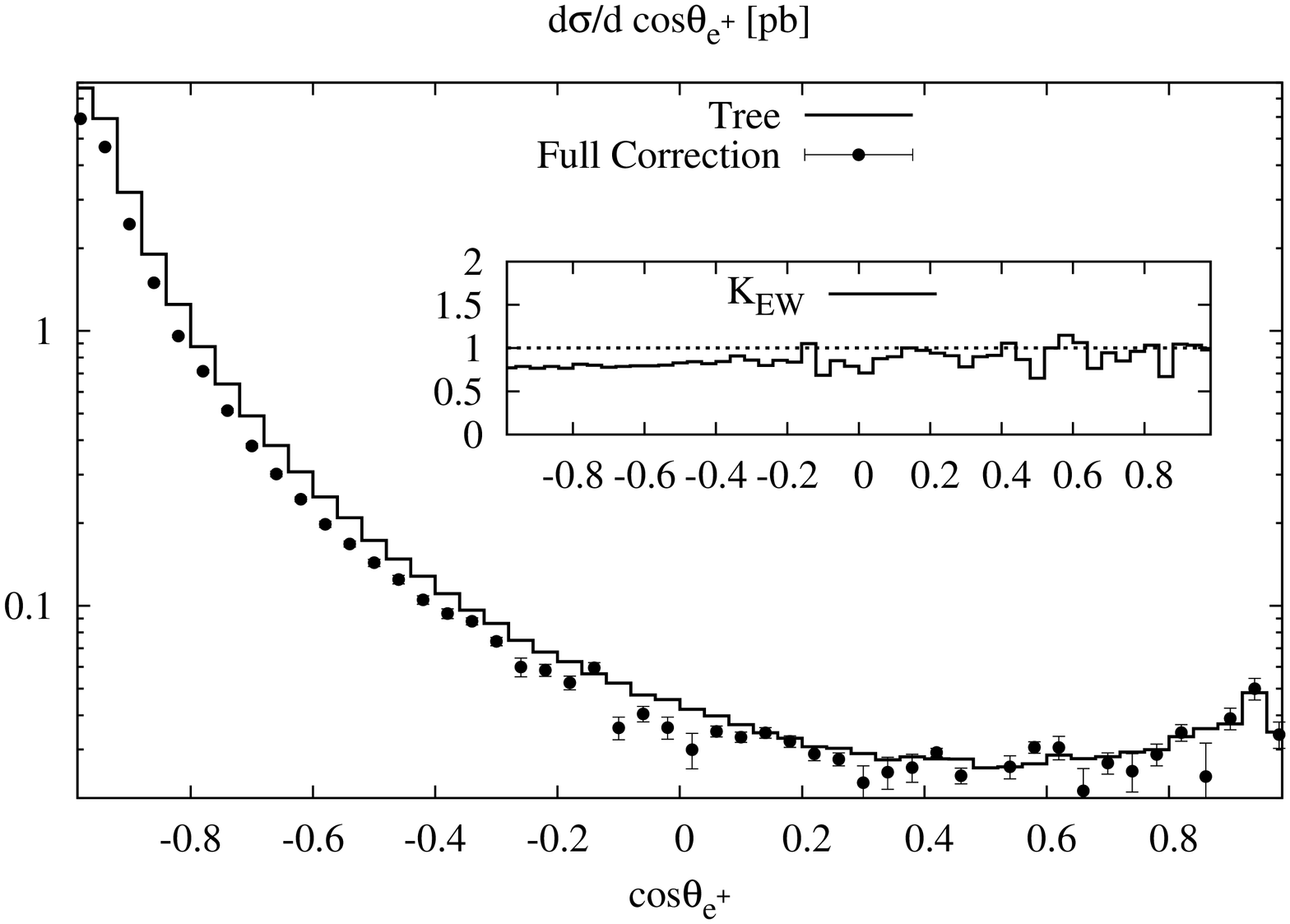}\\
\end{array}
$\end{center}
\caption{\label{e4} Differential cross sections as a function of the positron 
energy and cos$\theta_{e^+}$.
In the left (right) panel $\sqrt{s} = 250$ GeV ($\sqrt{s} = 1$ TeV).}
\end{figure}

Figure~\ref{e4} presents the differential cross sections as a function of 
the positron energy for $\sqrt{s}=250$ GeV and $\sqrt{s}=1$ TeV. The cross 
section increases with increasing positron energy. Two peaks appear in the 
distributions; the first of which is attributed to the highest-energy 
positron  $E_{e^+} \sim \dfrac{\sqrt{s}}{2}$ (or the smallest invariant 
mass of the photon and electron). The second peak corresponds to a 
minimum-energy photon emitted from the electron. This peak appears at 
$E_{e^+} \sim \dfrac{\sqrt{s}}{2} - E_{\gamma}^{\mathrm{min}} $. Within 
our resolution at $\sqrt{s}=1$ TeV, the two peaks overlap. The positron 
angular distributions in the final state are shown at $\sqrt{s}=250$ GeV 
and $\sqrt{s}=1$ TeV in the lower part of Fig.~\ref{e4}. Again, the 
radiative corrections make a sizeable impact.

\begin{figure}[hbtp]
\begin{center}$
\begin{array}{cc}
\hspace*{-0.6cm}\includegraphics[width=8.54cm,height=8.65cm, angle=-90]{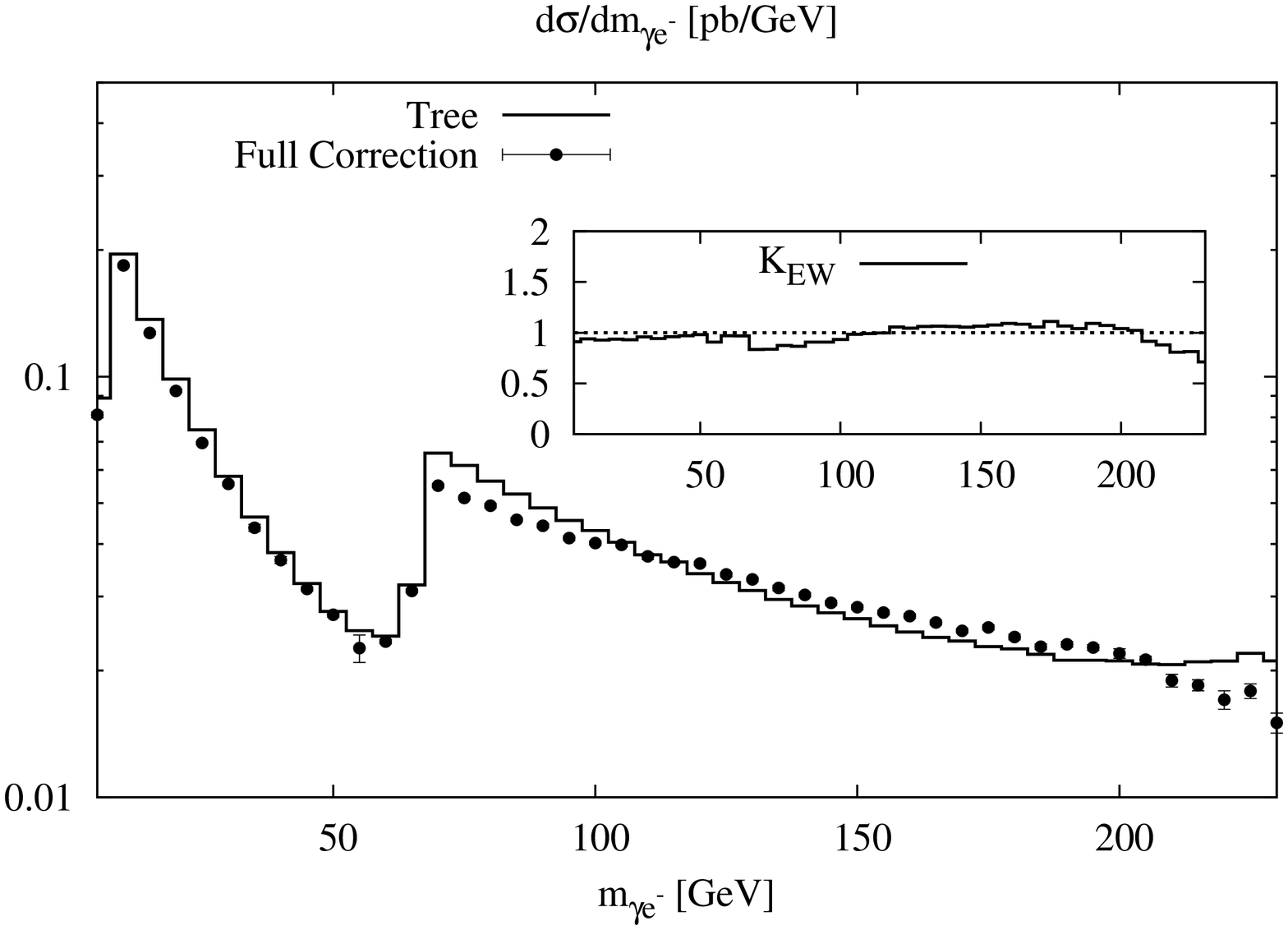} &
\hspace*{-1.0cm}\includegraphics[width=8.54cm,height=8.65cm, angle=-90]{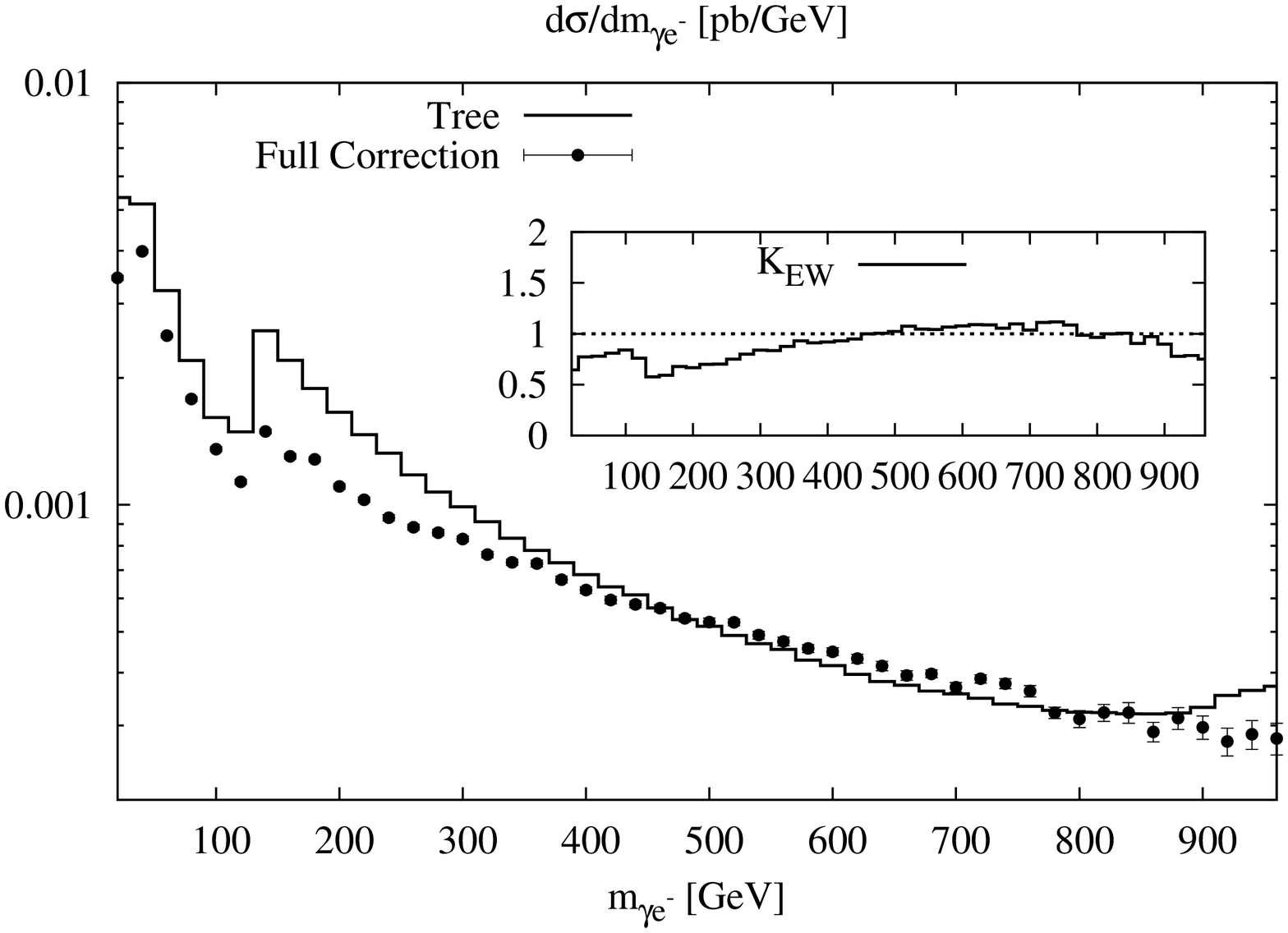}\\
\hspace{-0.27cm}\includegraphics[width=8.21cm,height=8.32cm, angle=-90]{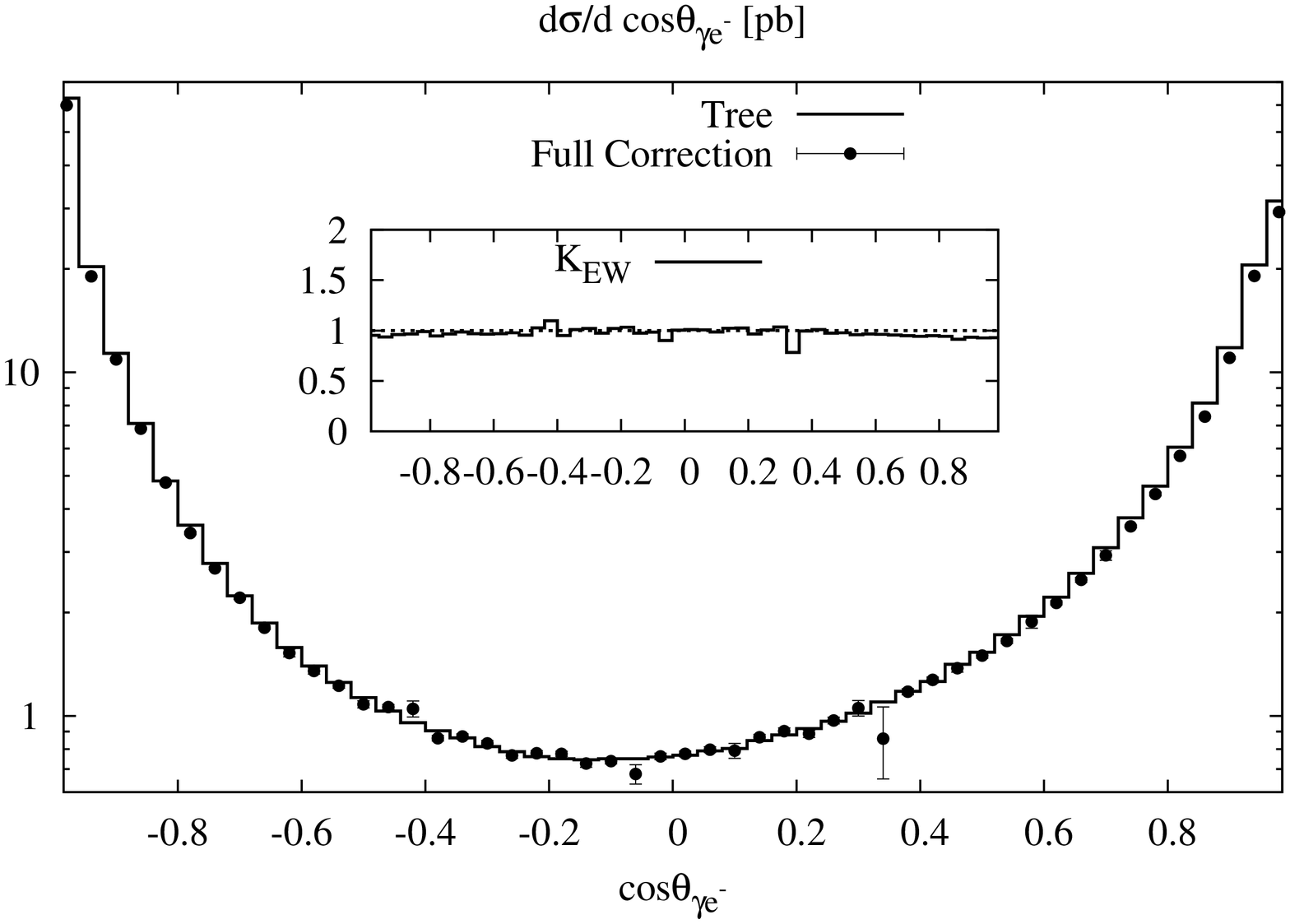} &
\hspace{-0.67cm}\includegraphics[width=8.21cm,height=8.32cm, angle=-90]{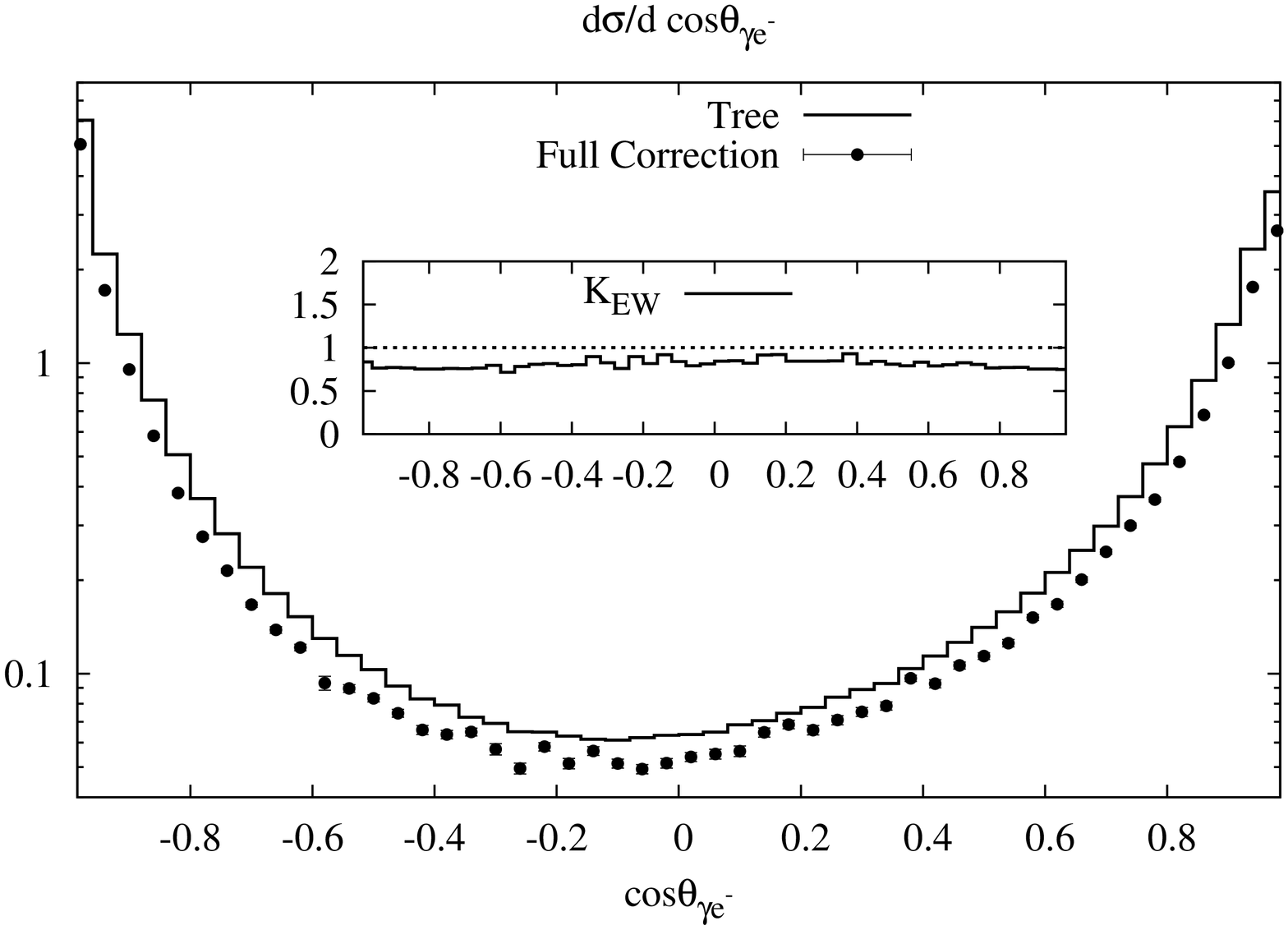}\\
\end{array}$
\end{center}
 \caption{\label{m34} Differential cross sections as a function of the 
invariant mass of the $e^-$ and the photon, $m_{\gamma e^-}$, and 
cos$\theta_{\gamma e^-}$. The left panels are for $\sqrt{s} = 250$ GeV and 
the right panels are for $\sqrt{s} = 1$ TeV.}
\end{figure}

A major concern for experiments is how to isolate the photon from the 
electron (and positron) in the final states because this would provide 
useful information for distinguishing $e^-e^+ \gamma$ from $e^-e^+$ events. 
Figure~\ref{m34} presents the distributions of the cross section as a 
function of the invariant mass of the $e^-$ and the photon ($m_{\gamma 
e^-}$) at $\sqrt{s}=250$ GeV and $\sqrt{s}=1$ TeV. The cross section 
decreases with increasing $m_{\gamma e^-}$. Two peaks appear in the 
distributions, having a similar origin as the peaks in the positron energy 
distributions. The lower part of Fig. \ref{m34} shows the angular 
distributions of the opening angle between the photon and the electron in 
the final state. The results indicate that the radiative corrections 
contribute significantly at the peaks and tails of the distributions. Thus, 
such corrections are important for distinguishing $e^-e^+ \gamma$ from 
$e^-e^+$ events.

\section{Conclusions}

Using the GRACE-Loop system, we calculated the full $\mathcal{O}(\alpha)$ 
electroweak radiative corrections to the $e^+e^- \rightarrow e^+e^-  
\gamma$ process for energies to be expected at the International Linear 
Collider.

The GRACE-Loop system incorporates a generalized nonlinear gauge-fixing 
condition that includes five gauge parameters. Combined with UV, IR 
finiteness and cutoff stability tests, they provides a powerful tool for 
testing the consistency of the results. The tests indicate that the 
numerical results are stable when quadruple precision is used.

We show that the full electroweak radiative corrections vary from $\sim 
-4\%$ to $\sim -21\%$ for a center-of-mass energy ranging from $250$ GeV to 
$1$ TeV. These corrections have a sizeable impact on the differential cross 
sections. Therefore, this calculation is important for determining the 
luminosity at the ILC.

In future work, we plan to incorporate the process with a soft 
bremsstrahlung photon and subsequently the full two-loop corrections to 
Bhabha scattering into the calculation.
 
\section*{Acknowledgments} We sincerely thank Prof. F.~Yuasa and Dr. 
N.~Watanabe for their valuable discussions and comments. The authors are 
grateful to Prof.~K. Tobimatsu and Prof.~M. Igarashi for useful discussions 
and their contributions. The work of T.U. was supported by the DFG through 
SFB/TR 9 ``Computational Particle Physics'' and the work of J.V. was 
supported by the ERC advanced grant 320651, ``HEPGAME.''

\section*{Appendix}
\label{numericalcheck}
The calculation is checked numerically at a point in phase space where the 
four components of the particles' momentum, $ p^{\mu }(p_x;p_y; p_z;  E)$, 
are
\begin{eqnarray}
 p_1^\mu & = & ( 0, \; 0, \;\ \  499.999999999738879960679048216325,\; \;500 )
                    \nonumber\\
 p_2^\mu & = & ( 0, \; 0, \;-499.999999999738879960679048216325, \; 500 )
                    \nonumber \\
 p_3^\mu & = & ( -103.078628242427254979669506380205, \nonumber \\ &&
     \; -114.633210803542408648432443344924, \nonumber \\ && 
     \; -471.180628984259439976275161034772, \nonumber \\ &&
     \;\ \  495.759177171207049330152475391965\; ) \nonumber  \\
 p_4^\mu & = & (\ \ \ \ 8.55713405427702202967532141788216, \nonumber \\ &&
     \;  -14.8872000707485148244530094120855, \nonumber \\ && 
     \;\ \; 72.9130813076746195344593973796190, \nonumber \\ &&
     \;\ \; 74.9077478983986818304861899869219\; ) \nonumber  \\
 p_5^\mu & = & (\ \ 94.5214941881502329499941849623225, \nonumber \\ &&
     \;\  129.520410874290923472885452757010, \nonumber \\ && 
     \;\  398.267547676584820441815763655153, \nonumber \\ &&
     \;\  429.333074930394268839361334621113\; ) \nonumber
\end{eqnarray}

The tables \ref{cuv}--\ref{gauge} below present the numerical results for 
the tests of the UV and IR finiteness and the gauge-parameter independence 
at this point in phase space. The results of the test of $k_c$-stability 
are presented in Table (\ref{kc}).

\begin{table}[htpb]
\centering
\begin{tabular}{l@{\hspace{1.5cm}}c} \hline 
$C_{UV}$       & $2\mathcal{\Re}(\mathcal{M}_{Loop}\mathcal{M}^+_{Tree})$ \\ \hline\hline
$0$            & $   -0.142224672059345022803237910656998      $ \\ \hline
$10^2$         & $   -0.142224672059345022803237910656997      $ \\ \hline
$10^4$         & $   -0.142224672059345022803237910657050      $ \\ \hline\hline
\end{tabular} \vspace{2mm}
\caption{\label{cuv} Test of independence of $C_{UV}$ with respect to 
amplitude. For the results given in this table, the nonlinear gauge 
parameters are $0$ and $\lambda=10^{-21}$ GeV, and we use $1$ TeV for the 
center-of-mass energy.}
\end{table}

\begin{table}[htpb]
\centering
\begin{tabular}{l@{\hspace{1.5cm}}c} \hline 
 $\lambda$ [GeV]           &  $2\mathcal{\Re}(\mathcal{M}_{Loop}\mathcal{M}^+_{Tree})$+ soft contribution  \\ \hline \hline 
 $10^{-21}$ 		    & $  -3.570620888259806801441498543829971 \cdot 10^{-2}$  \\ \hline 
 $10^{-25}$ 		    & $  -3.570620888259806801404094882895954 \cdot 10^{-2}$  \\ \hline 
 $10^{-30}$ 		    & $  -3.570620888259806801404090885240872 \cdot 10^{-2}$  \\ \hline \hline
\end{tabular} \vspace{2mm}
\caption{\label{lambda} Test of IR finiteness of amplitude. For the results 
given in this table, the nonlinear gauge parameters are $0$ and $C_{UV}=0$ 
and the center-of-mass energy is $1$ TeV.}
\end{table}
 
\begin{table}[htbp]
\centering
\begin{tabular}{l@{\hspace{1.5cm}}c} \hline 
 $(\tilde{\alpha}, \tilde{\beta}, \tilde{\delta},\tilde{\kappa}, \tilde{\epsilon} )  
 $   &  $2\mathcal{\Re}(\mathcal{M}_{Loop}\mathcal{M}^+_{Tree})$         \\ \hline \hline
(0, 0, 0, 0, 0)                  & $-0.142224672059345022803237910656998 $ \\ \hline
(10, 20, 30, 40 ,50)             & $-0.142224672059345022803237910657197 $ \\ \hline
(100,200,300,400,500)            & $-0.142224672059345022803237910505800 $ \\ \hline\hline
\end{tabular} \vspace{2mm}
\caption{\label{gauge} Gauge invariance of amplitude. For the results shown 
in this table, we set $C_{UV} = 0$, the photon mass is $10^{-21}$ GeV, and 
the center-of-mass energy is $1$ TeV.}
\end{table}

\begin{table}[htpb]
\centering
\begin{tabular}{c@{\hspace{1.5cm}}c@{\hspace{1.5cm}}c@{\hspace{1.5cm}}c} \hline 
 $k_c$ [GeV]     &$\sigma_{S}$ [pb]           & $\sigma_{H}$ [pb]         &$\sigma_{S+H}$ [pb]   \\ \hline \hline 
$10^{-3}$        &$ 7.873\pm 0.004$                    &$2.506\pm 0.002$                    &$10.379\pm 0.004$               \\ \hline
$10^{-2}$        &$ 8.401\pm 0.004$                    &$1.980\pm 0.001$                    &$10.381\pm 0.004$               \\ \hline
$10^{-1}$        &$ 8.932\pm 0.004$                    &$1.453\pm 0.001$                    &$10.385\pm 0.004$                \\ \hline\hline
\end{tabular} \vspace{2mm}
\caption{\label{kc} Test of $k_c$-stability. The photon mass is $10^{-21}$ 
GeV and the center-of-mass energy is $1$ TeV. The second column presents 
the soft-photon cross section and the third column presents the hard-photon 
cross section. The final column is the sum of both.}
\end{table}



\begin{thebibliography}{10}
\bibitem{Bozovic-Jelisavcic:2013aca}
  I.~Božović Jelisavčić, S.~Lukić, G.~Milutinović Dumbelović, M.~Pandurović and I.~Smiljanić,
  JINST {\bf 8} (2013) P08012
  [arXiv:1304.4082 [physics.acc-ph]].
  
\bibitem{tobimatsu1}
K.Tobimatsu and Y.Shimizu, 
Prog. Theor. Phys. {\bf 74} (1985), 567-575.
\bibitem{tobimatsu2}
K.Tobimatsu and Y.Shimizu, 
Prog. Theor. Phys. {\bf 75} (1986), 905-913.

\bibitem{Bohm:1984yt}
  M.~Bohm, A.~Denner, W.~Hollik and R.~Sommer,
  Phys.\ Lett.\ B {\bf 144} (1984) 414.

\bibitem{Bohm:1986fg}
  M.~Bohm, A.~Denner and W.~Hollik,
  Nucl.\ Phys.\ B {\bf 304} (1988) 687.
 
\bibitem{Berends:1987jm}
  F.~A.~Berends, R.~Kleiss and W.~Hollik,
  Nucl.\ Phys.\ B {\bf 304} (1988) 712.

\bibitem{Fleischer:2006ht}
  J.~Fleischer, J.~Gluza, A.~Lorca and T.~Riemann,
  Eur.\ J.\ Phys.\  {\bf 48} (2006) 35
  [hep-ph/0606210].

\bibitem{Penin:2005kf}
  A.~A.~Penin,
  Phys.\ Rev.\ Lett.\  {\bf 95} (2005) 010408
  [hep-ph/0501120].

\bibitem{Penin:2005eh}
  A.~A.~Penin,
  Nucl.\ Phys.\ B {\bf 734} (2006) 185
  [hep-ph/0508127].
  
\bibitem{Bonciani:2004gi}
  R.~Bonciani, A.~Ferroglia, P.~Mastrolia, E.~Remiddi and J.~J.~van der Bij,
  Nucl.\ Phys.\ B {\bf 701} (2004) 121
  [hep-ph/0405275].

\bibitem{Bonciani:2004qt}
  R.~Bonciani, A.~Ferroglia, P.~Mastrolia, E.~Remiddi and J.~J.~van der Bij,
  Nucl.\ Phys.\ B {\bf 716} (2005) 280
  [hep-ph/0411321].

\bibitem{Bonciani:2007eh}
  R.~Bonciani, A.~Ferroglia and A.~A.~Penin,
  Phys.\ Rev.\ Lett.\  {\bf 100} (2008) 131601
  [arXiv:0710.4775 [hep-ph]].

\bibitem{Bonciani:2008ep}
  R.~Bonciani, A.~Ferroglia and A.~A.~Penin,
  JHEP {\bf 0802} (2008) 080
  [arXiv:0802.2215 [hep-ph]].
  
\bibitem{Actis:2007fs}
  S.~Actis, M.~Czakon, J.~Gluza and T.~Riemann,
  Phys.\ Rev.\ Lett.\  {\bf 100} (2008) 131602
  [arXiv:0711.3847 [hep-ph]].
  
\bibitem{Actis:2008br}
  S.~Actis, M.~Czakon, J.~Gluza and T.~Riemann,
  Phys.\ Rev.\ D {\bf 78} (2008) 085019
  [arXiv:0807.4691 [hep-ph]].

  
\bibitem{Actis:2009uq}
  S.~Actis, P.~Mastrolia and G.~Ossola,
  Phys.\ Lett.\ B {\bf 682} (2010) 419
  [arXiv:0909.1750 [hep-ph]].
  
\bibitem{Penin:2011aa}
  A.~A.~Penin and G.~Ryan,
  JHEP {\bf 1111} (2011) 081
  [arXiv:1112.2171 [hep-ph]].
  
\bibitem{Tobimatsu:2001kb}
  K.~Tobimatsu and M.~Igarashi,
  Comput.\ Phys.\ Commun.\  {\bf 136} (2001) 105.
  
 \bibitem{Igarashi}
  M.~Igarashi et al,
  \textit{in preparation}.
  
\bibitem{Belanger:2003sd} 
  G.~Belanger, F.~Boudjema, J.~Fujimoto, T.~Ishikawa, T.~Kaneko, K.~Kato and Y.~Shimizu,
  Phys.\ Rept.\  {\bf 430}, 117 (2006)
  [hep-ph/0308080].

\bibitem{Belanger:2003ya}
  G.~Belanger, F.~Boudjema, J.~Fujimoto, T.~Ishikawa, T.~Kaneko, Y.~Kurihara, K.~Kato and Y.~Shimizu,
  Phys.\ Lett.\ B {\bf 576} (2003) 152
  [hep-ph/0309010].
\bibitem{Belanger:2003nm} 
  G.~Belanger, F.~Boudjema, J.~Fujimoto, T.~Ishikawa, T.~Kaneko, K.~Kato, Y.~Shimizu and Y.~Yasui,
  Phys.\ Lett.\ B {\bf 571}, 163 (2003)
  [hep-ph/0307029].

\bibitem{Belanger:2002me} 
  G.~Belanger, F.~Boudjema, J.~Fujimoto, T.~Ishikawa, T.~Kaneko, K.~Kato and Y.~Shimizu,
  Nucl.\ Phys.\ Proc.\ Suppl.\  {\bf 116}, 353 (2003)
  [hep-ph/0211268].

\bibitem{Zhang:2003jy}
  R.~-Y.~Zhang, W.~-G.~Ma, H.~Chen, Y.~-B.~Sun and H.~-S.~Hou,
  Phys.\ Lett.\ B {\bf 578} (2004) 349
  [hep-ph/0308203].

\bibitem{You:2003zq}
  Y.~You, W.~-G.~Ma, H.~Chen, R.~-Y.~Zhang, S.~Yan-Bin and H.~-S.~Hou,
  Phys.\ Lett.\ B {\bf 571} (2003) 85
  [hep-ph/0306036].

\bibitem{Denner:2003ri}
  A.~Denner, S.~Dittmaier, M.~Roth and M.~M.~Weber,
  Phys.\ Lett.\ B {\bf 575} (2003) 290
  [hep-ph/0307193].

\bibitem{Denner:2003zp}
  A.~Denner, S.~Dittmaier, M.~Roth and M.~M.~Weber,
  Nucl.\ Phys.\ B {\bf 680} (2004) 85
  [hep-ph/0309274].

\bibitem{Denner:2003yg}
  A.~Denner, S.~Dittmaier, M.~Roth and M.~M.~Weber,
  Phys.\ Lett.\ B {\bf 560} (2003) 196
  [hep-ph/0301189].

\bibitem{Denner:2003iy}
  A.~Denner, S.~Dittmaier, M.~Roth and M.~M.~Weber,
  Nucl.\ Phys.\ B {\bf 660} (2003) 289
  [hep-ph/0302198].

\bibitem{Kato:2005iw}
  K.~Kato, F.~Boudjema, J.~Fujimoto, T.~Ishikawa, T.~Kaneko, Y.~Kurihara, Y.~Shimizu and Y.~Yasui,
  PoS HEP {\bf 2005} (2006) 312.

\bibitem{kyotorc}
K.~Aoki, Z.~Hioki, R.~Kawabe, M.~Konuma and T.~Muta, Suppl. Prog.
Theor. Phys.
  {\bf 73} (1982) 1.
\bibitem{form}
J. A. M. Vermaseren: {\it New Features of FORM}; math-ph/0010025.
\bibitem{form4.0}
  J. Kuipers, T. Ueda, J.A.M. Vermaseren, J. Vollinga,
  {\em Comput.Phys.Commun.} {\bf 184(2-13)} 1453-1467
  
\bibitem{ff}
G. J. van Oldenborgh, {\em Comput. Phys. Commun.} {\bf 58} (1991)1.
\bibitem{looptools}
T. Hahn, {\tt LoopTools},
\verb+http://www.feynarts.de/looptools/+.

\bibitem{nlg-generalised}
F. Boudjema and E. Chopin, {\em Z.~Phys.} {\bf C73} (1996) 85;
hep-ph/9507396.

\bibitem{Khiem:2012bp}
  P.~H.~Khiem, J.~Fujimoto, T.~Ishikawa, T.~Kaneko, K.~Kato, Y.~Kurihara, Y.~Shimizu and T.~Ueda {\it et al.},
  Eur.\ Phys.\ J.\ C {\bf 73} (2013) 2400
  [arXiv:1211.1112 [hep-ph]].
\bibitem{supplement100}
J.~Fujimoto, M.~Igarashi, N.~Nakazawa, Y.~Shimizu and
K.~Tobimatsu, {\em Suppl.
Prog.~Theor.~Phys.} {\bf 100} (1990) 1.
\bibitem{bases}
S.~Kawabata, {\em Comp. Phys. Commun.} {\bf 41} (1986) 127; {\it
ibid.,} {\bf 88} (1995) 309.
\bibitem{mpi}
http://www.mcs.anl.gov/research/projects/mpi/
\bibitem{grace}
T. Ishikawa, T. Kaneko, K. Kato, S. Kawabata, Y. Shimizu and
H.~Tanaka, KEK
  Report 92-19, 1993, {\tt GRACE} manual Ver. 1.0.

\bibitem{Denner:2005nn}
A.~Denner and S.~Dittmaier,
Nucl.\ Phys.\ B {\bf 734} (2006) 62 [hep-ph/0509141].

\bibitem{Hioki:1995ex} 
  Z.~Hioki,
  Acta Phys.\ Polon.\ B {\bf 27}, 2573 (1996)
  [hep-ph/9510269].
\end{thebibliography}
\end{document}